\documentclass{elsart}
\journal{Physics Letters B}
\usepackage{cite}
\usepackage{amssymb}
\usepackage{subeqn}
\usepackage{psfig}
\def\simgt{\lower.7ex\hbox{$\;\stackrel{\textstyle>}{\sim}\;$}}
\def\simlt{\lower.7ex\hbox{$\;\stackrel{\textstyle<}{\sim}\;$}}
\begin{document}
\begin{frontmatter}
\begin{flushright}
IC/2000/74\\hep-ph/0006344\\ June 2000   
\end{flushright}
\title{Stable Q--balls from extra dimensions}  
\date{\today}
\author{D.\ A.\ Demir}
\address{The Abdus Salam International Centre for Theoretical Physics, Trieste, Italy}
\begin{abstract}
Given a bulk scalar field with sufficient self--interactions in a higher dimensional spacetime, 
it is shown that the continuous symmetries in four dimensions, induced by the topological 
structure of the compact manifold, naturally lead to formation of stable nontopological
solitons of Q--ball type. The mass per unit charge inside the soliton is bounded by the size 
of the extra dimensions, and it is thus stable with respect to decaying into excited 
levels of all bulk fields, irrespective of their bulk masses. A familiar example is the Standard Model 
in the bulk, where the Kaluza--Klein levels of the Higgs boson form a stable Q--ball. 
These stable solitons are natural candidates for the dark matter in the Universe.

\end{abstract}
\end{frontmatter}
\setcounter{footnote}{0} 

\section{Introduction}
In all theories of Kaluza--Klein (KK) type \cite{Overduin:1997pn} one hypothesizes that the spacetime is 
$4+\delta$ dimensional, and it has already a partially compactified structure ${\bf M}^{4}\times {\bf B}^{\delta}$ 
\cite{Scherk:1975fm}, instead of ${\bf M}^{4+\delta}$, where ${\bf B}^{\delta}$ is a compact $\delta$--dimensional 
space. The symmetries of ${\bf M}^{4}\times {\bf B}^{\delta}$ are the four--dimensional Poincare symmetries acting 
on ${\bf M}^{4}$ and some internal symmetry group generated by the topological structure of ${\bf B}^{\delta}$. The symmetries of
the internal manifold are observed as the symmetries of the four--dimensional lagrangian. For instance,
${\bf M}^{4}\times {\bf S}^{1}$ generates a $U(1)$ symmetry in four--dimensions whereas ${\bf M}^{4}\times {\bf S}^{2}$ realizes
an SU(2) symmetry \cite{Witten:1981me}. Unlike the common wisdom in KK theories where one
attempts to construct realistic gauge theories in four dimensions from a higher dimensional theory
with an appropriate internal manifold \cite{Witten:1981me}, here we consider a bulk field theory in
higher dimensions, and discuss possible effects of KK symmetries in the four--dimensional theory.

The extra dimensions could be spacelike or timelike. Compactification of the timelike   
coordinates cause probability and causality violation, and thus, phenomenologically 
they should not exceed Plackian sizes \cite{yndurain} unless there is some mechanism localizing
particles in the extra time coordinate \cite{gia}. On the other hand, within the present
experimental precision \cite{exp}, the compact spacelike dimensions could be as large as a $mm$
with important phenomenological implications \cite{nima}. Given a gauge theory in the bulk of
${\bf M}^{4}\times {\bf B}^{\delta}$ spacetime, the scalar sector of this theory must be such that
the gauge group should be spontaneously broken down to the subgroup required by physics   
in four dimensions. This is because of the fact that the KK compactification
cannot lead to symmetry breaking in lower dimensions. Hence, in what follows
we consider a bulk scalar field theory describing a physical spinless meson.

In what follows we will show that, irrespective of the symmetries and particle spectra of the
field theory in ${\bf M}^{4}\times {\bf B}^{\delta}$ bulk, the continuous symmetries of the four--dimensional
effective theory generated by the topological structure of ${\bf B}^{\delta}$ enable the formation of
nontopological solitons of Q--ball type \cite{Coleman:1985ki} if the scalar sector possesses
necessary interactions. This particularly necessitates that the internal manifold should be
compact ($e.g.$ there is no orbifolding), and at least the scalar fields should admit a
KK expansion ($e. g.$ they are not strictly localized on a topological defect \cite{topol,nima2}).

\section{Kaluza--Klein Q--Balls}

For simplicity we consider a $(1+4)$--dimensional spacetime with a partially compactified structure 
${\bf M}^{4}\times {\bf S}^1$ \cite{Scherk:1975fm}. Although we assume a 1--sphere for the internal space,
the results of the analysis will be general enough to cover higher dimensional compact spaces ${\bf B}^{\delta}$. 
As usual the metric tensor $G_{A B}(x_{\mu}, y)$ can be decomposed in terms of graviton $g_{\alpha \beta}(x_{\mu},y)$,  
graviphoton $A_{\alpha}(x_{\mu}, y)$ and graviscalar $\sigma(x_{\mu}, y)$ all living in the bulk 
\cite{Overduin:1997pn,brans}. However, all essential features of KK Q--balls can be understood by analyzing 
a self--interacting  bulk scalar field theory in a flat ${\bf M}^{4}\times {\bf S}^{1}$ background. Therefore, 
the KK excitations in the graviton sector, which are not essential for Q--ball formation, can be 
switched off by replacing the metric tensor with its vacuum expectation value \cite{duff} 
\begin{eqnarray}
\label{metric}
\langle G_{A B}(x_{\mu}, y) \rangle = \left(\begin{array}{cc}
 \eta_{\alpha \beta}& 0\\ 0&~ -1\end{array} \right)~.
\end{eqnarray}
Consequently, we consider a bulk scalar field $\Phi(x_{\mu}, y)$ with the action 
\begin{eqnarray}
\label{act5}
S_{5}=\int d^4 x d y \Big\{ \partial_{\mu} \Phi \partial^{\mu} \Phi -\left(\partial_{y}\Phi\right)^{2}-\overline{m}^2 \Phi^2 
-V(\Phi)\Big\}
\end{eqnarray}
where $\overline{m}$ is a bare mass parameter, and $V(\Phi)$ contains trilinear and higher order self--interaction terms. 

Letting $R$ be the radius of ${\bf S}^1$, the periodicity of $\Phi(x_\mu, y)$ in $y$ allows for the Fourier decomposition 
\begin{eqnarray}
\label{exp}
\Phi(x, y) = \sum_{n=-\infty}^{+\infty} \phi_{n}(x)\frac{e^{i n y/R}}{\sqrt{2 \pi R}}~.
\end{eqnarray}
where $\phi_{-n}(x)=\phi_{n}^{\ast}(x)$ due to the reality of $\Phi(x_{\mu}, y)$.
Hence, the theory in ${\bf M}^4$ now consists of
an infinite number of complex scalars
$\left\{\phi_n\right\}$
\begin{eqnarray}
\label{4dim}
S_{4}&=&\int d^{4} x \sum_{n}\Bigg[ \partial_{\mu} \phi_{n}\partial^{\mu} \phi_{-n}-\left(\overline{m}^2+ m_{n}^{2}\right)
\phi_{n}\phi_{-n} - V_{n}(\phi)\Bigg]
\end{eqnarray}
where $m_{n}=|n|/R$, and  $V_n(\phi)$ follows from $V(\Phi)$ after using (\ref{exp}) and integrating 
over  $y$. For instance, a  cubic interaction in $V(\Phi)$ contributes to $V_n(\phi)$ by $\sum_{k} \phi_n \phi_k
\phi_{-(n+k)}$. For the model under consideration, one can restrict the KK summations in
(\ref{4dim}) to $\mbox{[}0, \infty \mbox{]}$ interval by using the reality of $\Phi(x_\mu, y)$. However, 
when there are several scalar (not necessarily real) fields the notation of (\ref{4dim}) is appropriate.

As is clear from (\ref{4dim}), irrespective of what type of interactions and what
kind of fields participate in ${\bf M}^{4}\times {\bf S}^{1}$ theory, the lagrangian in ${\bf M}^4$ 
possesses a global $U(1)$ invariance generated by translations along ${\bf S}^{1}$:
\begin{eqnarray}
\label{u1kk}
U(1)_{KK}~:~\phi_{n}(x)\rightarrow \phi_{n}(x)~ e^{i n \alpha}~,
\end{eqnarray}
and all bulk fields in the theory ($e. g.$ fermions and gauge bosons) have the same 
transformation property as well. In particular, both KK charge and KK mass of 
the zero mode $\phi_0$ vanishes identically. In general, symmetry properties
of $\phi_{0}$ are identical to those of $\Phi(x,y)$. Consequently, in the
present analysis, it remains as a real scalar field.

By convention $V_n (0)=0$, and $\left\{\phi_n\right\}=0$ represents the ground 
state of the theory in which $U(1)_{KK}$ is unbroken. In this symmetric vacuum there are infinite number
of spinless mesons with masses $\sqrt{\overline{m}^2+ m_n^2}$. However, as was shown 
by Coleman  \cite{Coleman:1985ki}, extended objects appear in the spectrum if 
the total energy is minimized for fixed $U(1)_{KK}$ charge 
\begin{eqnarray} 
\label{char}
Q\equiv \sum_{n} \frac{n}{i} \int d^{3}\vec{x}\ \phi_{-n}(\vec{x},t)
\stackrel{\leftrightarrow}{\partial}_{t} \phi_{n}(\vec{x},t)~.
\end{eqnarray}
It is convenient to introduce a Lagrange multiplier $\omega$ to incorporate the constraint 
$Q\neq 0$ to the energy functional
\begin{eqnarray}
\label{ener}      
E_{\omega}&=&\omega Q +\int d^{3} \vec{x}\ \sum_{n}\Bigg[ \nabla \varphi_{-n}\nabla
\varphi_{n}+M_n^{2}(\omega)\ \varphi_{-n} \varphi_{n}+V_n(\varphi)\Bigg]
\end{eqnarray}
where $M_n^{2}(\omega)=\overline{m}^2+\left(1- \omega^2  R^2\right) m_n^2$. This expression 
for energy is to be minimized with respect to $\omega$ and $\varphi_n$, independently. 
That the energy functional is independent of time follows from the fact that the scalars $\phi_n$ 
rotate in the internal space with velocities proportional to their $U(1)_{KK}$ charge
\begin{eqnarray}
\phi_{n}(\vec{x},t)=\varphi_{n}(\vec{x})\ e^{i n \omega t}~.
\end{eqnarray}
From (\ref{ener}) it is clear that in the fixed KK charge sector the $n$--th KK level acquires 
a tachyonic mass $m_n^2  \omega^2  R^2$, which is what enables $\varphi_n$ to develop a nonvanishing
VEV in a domain of finite charge accumulation. To minimize (\ref{ener}) for fixed 
$\omega$ one has to solve an infinite number of coupled nonlinear differential equations \cite{bounce}
with necessarily spherically symmetric solutions \cite{sphersym}. Q--balls exist 
for small \cite{Kusenko:1997ad} (but large enough to allow for a semiclassical analysis) as well as large
\cite{Coleman:1985ki} $Q$ values. In the large $Q$ (thin--wall) limit the minimal
energy configuration is well approximated by a sphere of radius $R_Q$, such that $\varphi_n=\overline{\varphi}_n\neq 0$ 
for $|\vec{x}|\leq R_Q$, and $\varphi_n\equiv 0$ for $|\vec{x}| > R_Q$  \cite{Coleman:1985ki}. For this lump 
of matter to be stable with respect to decaying into a free scalar $\phi_n$ its total energy ($\equiv$ mass $M_{Q}$)
must be less than the total energy $N_n M_n$ of $N_n=Q/n$ scalars having total charge $Q$. Then the 
stability condition takes the form \cite{nonvanish}
\begin{eqnarray}
\label{cond}
\left(\frac{M_Q}{Q}\right)_{min} < \left(\frac{M_n(0)}{n}\right)_{min}~,
\end{eqnarray}
where $n$ refers to that KK state for which the right--hand side is a minimum. In this equation
the left hand--side is a function of the KK condensates $\overline{\varphi}_n$ 
\begin{eqnarray}
\label{mass}
\left(\frac{M_Q}{Q}\right)_{min}=\left(\frac{4\pi}{3} R_Q^{3}\ \frac{2 \omega}{Q}\ \sum_n
\left\{M_{n}^{2}(0)\ \overline{\varphi}_n \overline{\varphi}_{-n}+V_n(\overline{\varphi})\right\}\right)^{1/2},
\end{eqnarray}
and the right--hand side equals 
\begin{eqnarray}
\label{up}
\left(\frac{M_n (0)}{n}\right)_{min}=\frac{1}{R}
\end{eqnarray} 
independent of whether $\Phi(x_\mu,y)$ has a finite bare mass or not. Here the discussion has been restricted 
to thin--wall (large $Q$) regime in which $M_Q\sim Q/R$. In the thick--wall (small $Q$) limit $M_Q/Q$ receives
a smooth $Q$ dependence \cite{Kusenko:1997ad}. If the potential $V_n(\phi)$ is a flat one then 
$M_{Q}\propto Q^{3/4}$ rather than $Q$, and the value of the condensate inside the Q--ball can exceed 
the energy density by many orders of magnitude \cite{flat}.
 
The results obtained here can be generalized to more complicated scalar field theories such as
the supersymmetric theories with \cite{Boz:2000rn} or without \cite{Kusenko:1997zq} CP violation,
or possible extensions of the standard Higgs sector \cite{Demir:1999zi}. If the number of extra dimensions 
increase, say $\delta=2$ with a nontrivial ${\bf S}^{2}$ topology, then the four--dimensional theory has an SU(2) 
symmetry which admits Q--ball formation \cite{nonabel}. 

The stability condition (\ref{cond}) requires that the mass per unit charge inside the Q--ball is bounded
by the inverse compactification radius. According to the present experimental accuracy \cite{exp} 
$10^{-4}~{\rm eV} \simlt R^{-1} \simlt M_{Pl}$. Therefore, if the potential $V(\Phi)$ admits
a stable Q--ball solution then its mass per unit charge is constrained solely by the gravity experiments.

The zero mode $\phi_0$, which has neither KK mass nor KK charge, has the same Lagrangian and the same symmetries
as the bulk scalar $\Phi(x_\mu, y)$ apart from rescalings of the couplings. However, depending on the 
details of the potential $V(\Phi)$ it couples to higher KK levels. For instance, a cubic interaction of the form 
$\phi_0 \phi_n \phi_{-n}$ can give rise to a KK Q--ball similar to baryon or lepton balls in supersymmetry 
\cite{Kusenko:1997zq}. Since the exit or entrance of $\phi_0$ to Q--ball domain does not affect the total 
charge; it behaves as a self--interacting real scalar field for $|\vec{x}|\geq R_Q$, and becomes a component
of the Q--matter for $|\vec{x}|< R_Q$. 

The discussions above have concentrated on a bulk scalar field theory, in particular, 
the gravity sector has been ignored by adopting the vacuum configuration (\ref{metric}) 
for the metric tensor. However, a proper analysis must take into account the full
structure of the metric tensor $G_{A B}(x_\mu,y)$. Once this is done the particle
spectrum is widened by components of $G_{A B}(x_\mu,y)$, $i.e.$, the graviton 
$g_{\alpha \beta}(x_\mu,y)$, the graviphoton $A_{\alpha}(x_\mu, y)$, and the 
graviscalar $\sigma(x_\mu, y)$. The appearance of the graviphoton implies 
that the $U(1)_{KK}$ symmetry is now promoted to a local invariance. However, 
this does not invalidate the analysis above as the gauged $U(1)_{KK}$ also 
admits a Q--ball solution \cite{local}. The function of the graviscalar appears 
in the dilaton field $\overline{\sigma}\equiv \ln\{\sigma/(12 \sqrt{ \pi G_N}) \}$,
which itself is a candidate for the bulk scalar $\Phi(x_\mu, y)$ necessary for 
Q--ball formation. However, this requires the introduction self--interaction 
terms for $\overline{\sigma}(x_\mu, y)$ not following from the decomposition of 
$G_{AB}(x_\mu, y)$. After the compactification, like any matter field, all 
components of $G_{A B}(x_\mu,y)$ possess a $U(1)_{KK}$ invariance. Since the 
stability bound in (\ref{cond}) depends only on the KK masses and charges of
the bulk scalar (which could be $\overline{\sigma}(x_\mu, y)$ itself) practically 
all the results derived above remain valid.

\section{Electroweak Kaluza--Klein Q--balls}
In this section we investigate the formation and stability of the KK Q--balls within
bulk electroweak theory \cite{ignatios2}. Clearly the KK reduction cannot 
lead to symmetry breaking so that the standard model gauge symmetry must
be already broken in ${\bf M}^{4}\times {\bf S}^{1}$ bulk. Therefore, the bulk Higgs 
boson field $h(x_\mu, y)$ replaces the bulk scalar $\Phi(x_\mu, y)$, and 
its self--inetraction Lagrangian provides the necessary potential 
\begin{eqnarray}
\label{smpot}
V(h)= \overline{m}^{2}\ h^{2} + \overline{\lambda}\overline{v} h^{3} + \frac{1}{4}\overline{\lambda} h^4
\end{eqnarray}
where $\overline{v}^2= \overline{m}^{2}/\overline{\lambda} \approx \left(246.2~{\rm GeV}\right)^{2}/\left(2 \pi R\right)$. Since 
there exists a nonvanishing cubic coupling, there is necessarily a Q--ball solution 
\cite{nonvanish,Coleman:1985ki,Demir:1999zi}. As the bulk Higgs scalar  $h(x_\mu, y)$ couples to 
all standard model fields so does the resulting KK Q--ball. This then raises the question of stability
with respect to decaying into the standard model spectrum. However, once a stable Q--ball solution is 
formed for (\ref{smpot}) it is then clear that the disassociation of the Q--matter to any field, 
irrespective of its bulk mass, is forbidden kinematically. This follows from the stability 
bound (\ref{cond}), which states that the mass per unit charge inside the Q--ball is less than two--particle 
threshold. To illustrate this point consider a bulk fermion $\Psi(x_\mu, y)$ which may have a 
mass $m_\psi$. If this mass is not generated via the spontaneous symmetry breaking then $\Psi(x_\mu, y)$ 
does not couple to $h(x_\mu, y)$, and thus, there is no question of stability. On the other hand,
if $m_{\psi}$ is proportional to $\overline{v}$ then Q--ball would evaporate \cite{evaporate} 
by emitting $\overline{\psi_n}\psi_{n}$ pairs from the surface were not it for the stability condition 
(\ref{cond}). Indeed,  $n$--th KK level has a mass $|n|/R$ so that kinematically the KK Q--ball cannot 
decay to charged fermion pairs according to (\ref{cond}). This absolute stability property is 
exclusive to KK Q--balls in which the zero modes are devoid of KK charge, and mass per unit charge 
inside the Q--ball is bounded by the inverse compactification radius. For comparison, one may recall the
lepton balls of supersymmetry \cite{Kusenko:1997zq} which necessarily evaporate by emitting 
neutrinos \cite{evaporate} from the surface. Irrespective of whether $U(1)_{KK}$ is global \cite{Coleman:1985ki}
or local \cite{local} there is a Q--ball solution, and therefore, inclusion of gravity sector
does not affect the stability of the Higgs KK Q--balls.

The discussion above would be sufficient for having stable Higgs KK Q--balls were not it 
for the absence of chiral fermions. Indeed, the KK reduction produces equal numbers of left--
and right--handed fermions so that the effective four--dimensional theory does not possess a 
definite chirality. In fact, to have chiral fermions in four dimensions it is necessary to 
perform an orbifolding of the extra dimension, and this, however, breaks the 
$U(1)_{KK}$ symmetry down to its $Z_2$ subgroup, leaving no continuous symmetry to develop Q--balls. 

However, orbifolding is not the only way of inducing chiral fermions in ${\bf M}^{4}$. Indeed,
if the dynamical theory in ${\bf M}^{5}$ possesses multiple discrete degenerate vacua then 
there is always a massless chiral fermion \cite{index} localized on the topological defect 
interpolating among different vacuum domains \cite{topol}. To be specific, let 
$\Theta(x_\mu, y)$ be a bulk scalar field in ${\bf M}^{5}$ with two degerate vacua, 
then there arises a wall profile $\Theta_{wall}(y)$ which interpolates between the 
vacua situated at $y\rightarrow \pm \infty$. In the potential induced by $\Theta_{wall}(y)$,
the zero mode fermion possesses a definite chirality and remains stuck to the 
wall \cite{index,topol} whereas the excited levels are all paired to have no 
preferred chirality \cite{nima2}. 

The localization mechanism above is valid for ${\bf M}^{5}$ where none of the
dimensions is compact. For spacetimes with a compact coordinate
$e.g.$ ${\bf M}^{4}\times {\bf S}^{1}$, besides  appropriate modifications 
in the underlying theory in order to allow for the formation of the walls, 
one must require $M_{wall} > > R^{-1}$ where $M_{wall}$ is the characteristic 
mass scale of the wall--former, $\Theta(x_\mu, y)$. The postion of the 
wall along ${\bf S}^{1}$ may have a slight $x_\mu$ dependence;  $y_0=y_{0}(x_\mu)$ \cite{shif,shif2}.
However, since the translational symmetry along ${\bf S}^{1}$ is spontaneously
broken, this position is equivalent to the static solution $\Theta_{wall}(y)$ 
as $y_{0}(x_\mu)$ will be swallowed by the graviphoton $A_{\mu}(x)$ to acquire
a mass. To have a nontrivial tilting of the wall with respect to ${\bf S}^{1}$
the graviphoton should be projected out \cite{shif}. For this purpose, it is convenient 
to impose a global $U(1)_I$ invariance on the system under which the wall--former 
transforms as $\Theta(x_\mu, y) \rightarrow e^{i q_I \alpha} \Theta(x_\mu, y)$.
In addition to this, there is a local $U(1)_{KK}$ invariance defined in (\ref{u1kk}).
These two symmetries are, however, spontaneously broken down to a global 
$\widetilde{U}(1)_{KK}\equiv \sin\alpha\ U(1)_{I} + \cos\alpha\ U(1)_{KK}$ 
invariance once $\Theta_{n}$ develops a VEV for a nonvanishing $n$ where 
$|n|\in (0, M_{wall} R)$ \cite{shif,shif2}. Any $n$ in this interval can be chosen 
to be the vacuum configuration, as the mapping between the vacuum circle and 
${\bf S}^{1}$ yields a topologically stable configuration. The remnant $\widetilde{U}(1)_{KK}$
invariance, however, can generate a KK Q--ball if the potential of the wall--former
has enough interactions. Practically, the most important change arises in the 
stability condition (\ref{cond}) where now mass per unit charge inside the 
Q--ball is bounded by $|\cos \alpha|/R$ intead of $1/R$. Therefore, the breaking 
direction in $U(1)_I\times U(1)_{KK}$ space, $\alpha$, enters the game. All
the standard model particles, which are the zero modes of the corresponding 
fields in ${\bf M}^{4}\times {\bf S}^{1}$, are localized on the wall
staying on ${\bf S}^{1}$. The level splittings ($\sim M_{wall}$ \cite{topol,nima2}) 
of their excited modes are much more coarser than the KK splitting 
($\sim R^{-1}$). Ignoring the former (which would be exact in the limit
when the wall shrinks to a $\delta$--function) the latter enables the 
formation of the KK Q--ball. In such a picture the wall--former plays 
two r{\^o}les: ($i$) it traps the massless quarks, leptons, Higgs,
and gauge fields \cite{topol} whereby reproducing the four--dimensional 
world, and ($ii$) its KK modes with global $\widetilde{U}(1)_{KK}$ symmetry
form a Q--ball of the type studied above. Since the tilt angle $\alpha$ determies the
$\widetilde{U}(1)_{KK}$ charge of every field , the kinematic stability of the 
Q--matter with respect to decaying into other particles still persists.

\section{Conclusion}

In this work we have shown that a bulk scalar theory with sufficient interactions
naturally accomodate nontopological solitons of Q--ball type after compactifying
over the extra dimensions. The mass per unit charge inside the soliton is bounded
by the inverse compactification radius; hence, it is absolutely stable with respect
to decaying into the KK towers of all particles in the theory, including the
gravity sector. A more realistic setting involving chiral fermions with the inclusion
of the gravity sector arises approximately within the tilted wall scenario of \cite{shif,shif2}.

The fact that the KK Q--balls are stable makes them a perfect candidate for the dark matter in the universe. Apart from
electroweak \cite{Demir:1999zi} or GMSB \cite{gmsb} Q--balls, in general, baryon
or lepton balls suffer from evaporation \cite{evaporate}. In supersymmetric theories, 
for instance, it is necessary to produce large enough baryon or lepton balls \cite{flat} 
to have them survive up to the present epoch \cite{dark}.

It is clear that, formation of the KK Q--balls is effectively a decompactification process. 
This follows from the fact that, at least for the bulk scalar sector, 
all excited KK modes are imprisoned into a finite spatial domain which contains 
a fixed KK charge. The topology of the extra dimensions determines the symmetries of
the four--dimensional effective theory, and minimization of the energy 
for fixed conserved charge maps all KK modes into a nontopological stable 
soliton. The ultimate answer to the existence or absence of the Q--balls (the KK Q--balls in particular)
will come from the present and future experiments \cite{Q-ball-exp}.

\vspace{0.5cm}
The author is grateful to Gia Dvali for many helpful discussions and suggestions,
and for a careful reading of the manuscript. Also it is a pleasure to thank E. Caceres,
R. Hernandez, A. \" Ozpineci and K. Ray for fruitful discussions.

\end{document}